\documentclass[10pt,twocolumn,journal,tse]{IEEEtran}
\usepackage{amsmath, mathtools, amssymb, amsthm, amsfonts}
\usepackage[numbers,comma,sort&compress]{natbib}
\usepackage{graphicx}
\usepackage{color}
\usepackage{caption}
\usepackage{subcaption}
\usepackage{float}
\usepackage{placeins}
\usepackage{listings}
\usepackage{multicol}
\usepackage{bicaption}
\usepackage{array}
\usepackage{cite}
\usepackage{rotating}

\usepackage{lipsum}
\usepackage{graphicx}
\usepackage{url}
\usepackage{authblk}

\date{}
\title{An Improved Energy-Aware Clustering Method for the Regional Queries in the Internet of Things}
\author[1,*]{ Arezoo  Khatibi\thanks{Arezoo  Khatibi, correspond email: \protect\url{arezoo.khatibi@grad.kashanu.ac.ir}, Faculty of Computer Science,  University of Kashan,  BLVD Ghotb Ravandi 6 kilometers, Kashan Iran.}}
\author[2]{Omid Khatibi}
\affil[1]{Faculty of Computer Science, University of Kashan, Kashan, Iran}
\affil[2]{Faculty of Mathematics, University of Vienna,Vienna, Austria}
\begin{document}
\maketitle

\begin{abstract}
 We will offer a method to improve energy efficient consumption for processing queries on the Internet of Things. We focused on an energy efficient hierarchical clustering index tree such that we can facilitate time-correlated region queries in the I.O.T (Internet of Things). We try to improve clustering and make a change on its proposed index tree. We try to do this by optimizing the query processing.  We improve clustering to increase the accuracy of the Internet of Things and prevent the network from disconnecting. In the article that we have chosen, there is a heterogeneous cluster which means there exists a large data difference in the two ends of a cluster. Also, it often happens that the same information is sent to the base station by two overlapping clusters; therefore, we save energy by eliminating duplicated data.

\end{abstract}
\begin{IEEEkeywords}
Keywords: Internet of Things, Time-dependent query processing, Hierarchical clustering.
\end{IEEEkeywords}
\section{Introduction}
\IEEEPARstart{A}{ccess} remote control of everything in the world  will intelligently be provided through the internet in the next 20 years. We will offer a method to improve energy efficient consumption for processing queries on the Internet of things. The importance of the issue becomes clear when, for example, there is a volcanic outbreak  existing  and outdated sensors inside  do not work because the power contained in its sensors has deleted and therefore could not post the correct data as notified; so, we collected data from the last few days. They are deprived and cannot afford essential tasks before the occurrence of the accident. Suppose you want to carry the blood platelets from the blood transfusion center to the hospital and during transfer, the temperature of the enclosure inside the car rises [1] rendering the platelets useless. It could corrupt since the necessary information about the high temperature was not reported to the base station in time, which could lead to a patient's death after an injection by using the corrupt product. Assume there are sensors in rooms of babies that are not able to explain the pain. That is connected to a pain assessment center, and we have the ability to send baby face videos by using the Internet of Things; hence, we can diagnose pain. The energy sensor in this case had finished and a newborn could have been in the first stage of a disease such as cataracts diagnosed too late leading the illness  to spread and cause blindness [2]. Suppose a sock for the prevention of diabetic ulcers is invented  where sensors are used to recognize increase in pressure in one area of the foot and use SMS to notify the patient via text. The patients expect the sensors to work properly. Thus, the sensors should not run out of, an energy loss could lead to a lost leg. We have important materials in a warehouse outside of town where we must continuously monitor and maintain the temperature and humidity of the environment, and we do so using the Internet of Things. The power of the sensors is finished when a robber enters the store and is thus not able to notify the police. These examples show that energy efficient consumption in wireless sensor networks, which are a large part of the Internet of Things is very important. The 'thing' is a device that has several sensors. When these things connect together through the internet, the Internet of Things creates an id and a specific IP [3]. One example of the Internet of Things is a Pain evaluation System. In which baby's room is networked, a video records the baby's facial expressions, and then is sent as a data stream  to the nerve center of the hospital. This method can help the families to assess whether the baby is crying out of pain or  another reason. This  method is useful for newborns and disabled persons who are unable to talk. The structure of the article is as follows: We will first explain the Internet of Things and then examine the methods of clustering on the Internet of Things. Since most of the Internet of Things consists of sensor nodes, this chapter is very similar to clustering in a wireless sensor network. Then we will describe our proposed method and then we will conclude with the results of our experiments.
\section{related works}
The database management system operates on the data stored in databases and the data flow management system operates on the data that must be respond to queries related to this data and generated by sensors at a given time interval. Tang et al [10] manage the data flow on their queries and split the area of wireless sensors into grid cells, and then propose a hierarchy clustering energy saving index tree for the grid cells. They created a time-dependent query technique for giving continuous queries.
Initially, sensor nodes report their values to the base station. If these values change significantly in comparison to the previously reported values,
they report their values to the base station again.
 The queries are processed by collecting the values of the sensors stored at the base station and analyzing them. They reduce energy consumption and provide an energy-saving hierarchy clustering index tree to facilitate time-sensitive regional querying on the Internet of Things. Tang et al [10] represented an energy model focused on building a routing tree, finding cluster centers was not important to them. They found the weights of the cells and the stages of their proposed tree construction are as follows: Making the shortest route to the base station. In [10], four algorithms were presented in which the first two algorithms comprise the grid cells, the weigh of the grid cells was then calculated, and the hierarchy tree was drawn.
On the other hand, [10] didn't consider overlapping clusters. It often happens that some geographic coordinates‏‫‬‬‬ are considered by two or more clusters. In fact, repeated data is taken and sent to the base station. If these geographic coordinates are considered by only one cluster, it will save energy. We proved data deletion saves energy.
So far, there are some methods for choosing the right cluster head within a cluster for wireless sensor networks. Since most elements of the Internet of Things use wireless sensor networks, these methods can be used to select the proper cluster head within a cluster in the Internet of Things. We will give a brief explanation of two  methods in this regard. Abdoulsalam and Kamer in 2010 proposed a method called W-Leach. In W-Leach protocol  a cluster head was selected based on the weight of the sensors and the network weights were based on sensor densities and their residual energy. The SNs density is determined by the number of live sensors in a specified region divided by the total number of live sensors in the network.
 The sensitivity and density of each sensor are based on the following 1- greater density $ \Rightarrow $ more weight 2- more energy remaining $ \Rightarrow $ more weight. If no sensor was found within the range,  the density is 1. If all the sensors are within the transmission range then the density is set to 'n'. In order to send data to the cluster head, sensors are selected whose densities are lower than the predefined threshold and compared to sensors that have a higher density in a network. Otherwise, they will wait for the next round to transfer the data to the corresponding cluster heads. Farooq in 2010 proposed a method called Mr.-Leach. It divides the network into different layers of clusters.  Mr.-Leach applies the concept of equal clustering which every node in the layer gets to the base station with the same number of hopes. It divides the network into different layers of clusters. Choosing cluster heads and other substrings at the second level is carried out by the base station. Therefore, the cost of calculations at the sensor level can be reduced. This is done in three phases: Creating a cluster in low level, cluster recognition at different levels by the base station, and scheduling. Pattem S extended " the impact of spatial correlation or routing with compression in wireless sensor networks [5]". In sensor networks, "low latency and an energy efficient routing tree for wireless sensor networks with multiple mobile sinks" is established by Han S-W [4]. Tyagi S and Neeraj Kumar made a full study of "A systematic review on clustering and routing techniques based upon leach protocol for wireless sensor networks"[6]. Gastpar et al [7] found the repetitive messages that the sensor nodes had seen in relation to the relationship between messages that needed an environment for storing data and a technique for finding duplicate data that caused network latency and costly network implementation.  In [8], Al-Turjmana FM expanded "Quantifying connectivity in wireless sensor networks with grid-based deployments."Peng I-H and Chen Y-W the process by which "Energy consumption bounds analysis and its applications for grid-based wireless sensor networks" has attracted considerable attention recently [9]. Pathan A-SK used "A secure energy-efficient routing protocol for WSN"[11]. They made the routing tree that the query was transmitted to the child node and the answer, in the opposite direction, returned to its root, which needed to know the parent-child relationship and had high energy consumption. Soheili and Kalogeraki spoke about" Spatial queries in sensor networks "in the proceeding of the 13th annual ACM international workshop on geographic information systems [12]." They reduced the number of nodes in a query, but this needed to know the parent-child relationship and had a high energy consumption.  Caione, Brunelli and Benini implemented and investigated "Distributed compressive sampling for lifetime optimization in dense wireless sensor networks[13]." Sardari, Beirami, Zou, Fekri  proposed " Content-aware network data compression by using joint memorization and clustering " [14]. The advantage of some methods of WSN is visible in [15] and [16]. Tan, Korpeoglu and Stojmenovic computed power-efficient data aggregation trees for sensor networks [17]. By using Service-Oriented Architecture, Guinard, Trifa, Karnouskos, Spiess, and Savio developed a process to run instances of real-world services, select and  dynamically query [18]. Kang, Lee, Kim, Choi, Im, and Kang E-Y utilized In-network query for wireless sensor networks [19].
\begin{table}[htbp]
  \centering
  \caption{ : Comparison of 10 methods of determining of  cluster head on the Internet of Things [6].\label{table:1}}
  \begin{small}
  \tiny
\begin{tabular}{|p{1.16cm}|p{0.22cm}|p{0.35cm}|p{0.50cm}|p{0.57cm}|p{0.45cm}|p{0.45cm}|p{0.94cm}|p{0.40cm}|p{0.40cm}} 
  \hline
       &Delay&Energy saving&load balance&Scalability&fault tolerance&Moving Nodes&Communication cost&How to transfer data\\ \hline
    Handy in 2002 & Low  & Good   & Not tested& Low &Not tested&Nodes are fixed&Top&Single hop\\ \hline
    Nego in 2007&Much&Very good&Not tested&too much&Not tested&Nodes are quasi constants&Not tested&Not tested\\ \hline
    Yang in 2010 & Not tested & Good   & Not tested& Not tested &Not tested&Nodes are fixed&Not tested&Single hop\\ \hline
    Wang in 2010 & Not tested & Very good  & Not tested& Not tested &Not tested&Nodes are fixed&Not tested&Single hop\\ \hline
    Sen in 2011 & Not tested  &Good  &Yes& Not tested &Not tested&Nodes are fixed&Not tested&Single hop\\ \hline
    Duan in 2009 & Not tested  & Good   & Yes& Not tested &Not tested&Nodes are fixed&Top&Single hop\\ \hline
    WALFA in 2009 & Not tested & Good   & Yes&much &Not tested&Nodes are fixed&Not tested&Multi- hops\\ \hline
    FAROOQ in 2010 & Not tested & Very good   & Yes& Not tested &Not tested&Nodes are fixed&Not tested&Multi- hops\\ \hline
    ABDOULSALAM and KAMER in 2010 &Not tested  & Very good  & Not tested& Not tested &Not tested&Nodes are fixed&Not tested&Single hop\\ \hline
    JUT in 2011 & Not tested & Good   & Yes& Not tested &Not tested&Nodes are fixed&Not tested&Single hop\\ \hline
  \end{tabular}
  
\end{small}
\end{table}

 As a matter of fact, for the minimum spanning tree and the shortest past tree approaches, we need energy efficient correlated data aggregation for wireless sensor networks [20]. Park S-J et al used a spatial index tree and broadcast queries that had high energy consumption.

In the Internet of Things, smart things communicate with each other. (Of course, it's going to happen in the next 20 years). To satisfy the specific requirements of all users when a zone is to be monitored continuously, the strategy to be implemented is to sense and independently collect the data for each query of the sub-areas. This method is not energy efficient because values reported by sensors may be the same as values sensed during the previous time interval.
In [10], there was no attention given to the problem of filtering data, though data filtering is a very effective way to reduce energy consumption.  If every sensor sends its data directly to the base station, significant amounts of memory will be needed and, of course, the accuracy of the data will increase. Tang et al [10] tried to improve this method, which consumes a lot of energy.
\begin{figure*}[htbp!]
 \centering
        \includegraphics[width=7in,height=12.5cm]{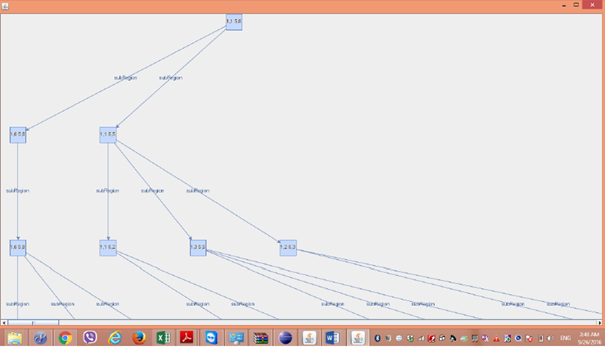}
        \centering
        \caption{"Results of the implementation of algorithms of reference[7] graphically in the Java environment."}
        \label{Graph A}
        
\end{figure*}
\section{Our solution}
 Their method was that when a major modification happened in the sensed data by a sensor then it would be reported to the base station. They used a clustering method in which the size of the cluster was not important and some clusters might be large while others
could be small.
In our proposed approach, we consider data filtered meaning the data will be filtered by the header in a grid cell. Functions that are used when data is filtered are typically maximizing, minimizing, or averaging on a header node, and we focused on the averaging function in our scenario design as well as the implementation of our proposed approach.

 For example, if data filtering that is executed by a header in a grid cell is an averaging function and the cluster is greater than a certain limit, the data collected at the lowest point by a node has great difference with the data that lies at the highest point of the cluster. We suggest adding a few headers to this cluster and, in fact, breaking this cluster and creating new clusters based on the K-MEANS algorithm.

Also, in the first and second algorithms given in [10], header selection inside a grid cell is based on the Leach method. According to Table 1 on the previous page, we offered a header selection  based on the Abdoulsalam $\&$ Kamer method (in 2010) since this method implements hierarchical clustering and energy consumption is reduced substantially. A method can be considered in accordance with Table 1 if it uses load balancing and significantly reduces energy consumption like Mr. Farooq's method (in 2010). On the other hand, when a cluster is large, the number of leaves that are children of a higher level in the index tree provided by Tang et al [10] increases too much, and transmitting by only one parent to a higher level especially during the use of averaging method means that an accurate filter of data has not been performed.
\begin{figure*}[htbp!]
 \centering
        \includegraphics[width=5in,height=10.5cm]{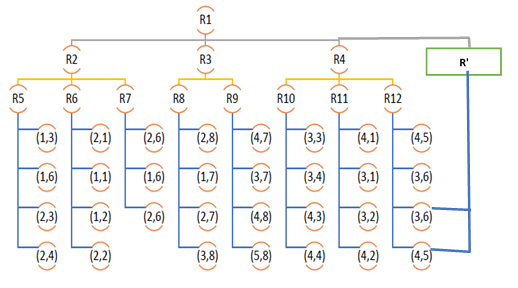}
        \centering
        \caption{" Proposed index tree based on the simulation of wireless sensor nodes of reference[7]".}
        \label{Graph B}
        
\end{figure*}
For example, in one scenario we want to make temperature averaging, the information of a leaf that is a grid cell may show a temperature of 2 degrees and the temperature of its neighbor leaf is slightly  different from this temperature. But the leaf is farther than the leaf that we are concentrating on and it has a  common father with our desired leaf at a higher level, his temperature is 22 degrees. In the averaging, the header that appears as the parent of these leaves sends a temperature of 12 degrees to the root of the tree or the base station. In these situations,  my suggestion is that when decisions and analysis are made by the base station to reduce  error, a subtree must be used that sends leaf information directly to the root of the tree. We show our proposed index tree in Figure 2 and we propose adding two functions to the algorithm 1 and 2 in [10], which are used to construct the index tree. 1- Whether the cluster is broken or not? This is done by considering an upper threshold value for the number of nodes in a cluster. For example, if the number of nodes exceeds a certain limit, the cluster will be broken.
2- Add the routine of constructing a sub-tree in a large cluster that is broken, the leaves are the grid cells and its root is the header cluster, which directly connects to the root of the tree. We have written a program in C++ that gets coordinates information including their temperature, from an input file, as well as the number of minimum possible clusters and maximum possible clusters. These items were named as m and M in the algorithms of [10]. Then, we will sort them and build a hierarchical tree. At the same time, the average values of the leaves are calculated in  the father sensor.
 Then we receive a query from the user and the program determines to which regions the query is related and returns the answer. We changed the number of clusters shown with m and M, and conclude that when the number of clusters increases, the query response is closer to the actual data value of a sensor. We implemented the algorithms 1, 2, 3, and 4 in the paper [10] graphically in the Java environment and by changing the value of m, we increased the depth of the tree. We found that if the depth of the tree increased, it would be better. In this status, the average data in the header is closer to the real data of the sensor. This program first creates grid cells and then calculates the weight of the grid cells. It then draws the corresponding hierarchical tree. The results of this program are presented graphically in Figure 1. Our environment can be a closed environment or a large forest or underwater data. The data in our system can be the wind blowing or temperature or humidity. Data sensed with sensors is sent to the header in the related grid cell and filtered by this header. There are several functions for filtering data, which in our scenario we use the average function to filter data. In [10], the clusters are asymmetric and clustering is done based on proximity. 
 \begin{figure*}[htbp]
 \centering
        \includegraphics[width=4in,height=9.5cm]{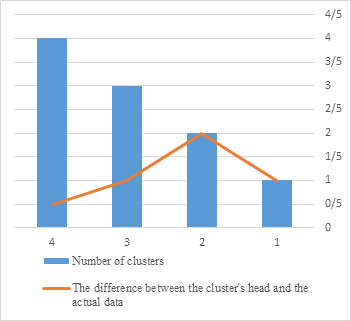}
        \centering
        \caption{ Comparison the difference between the header data and the actual data of one node.}
        \label{Graph C}

\end{figure*}
 Physically, the sensor's data is close to its next sensor, but ultimately the sensor located at the lowest point of the cluster has large difference in data to a sensor that is placed at the highest point of the cluster. It is wrong that only an average of this big cluster is to be reported to the base station, and another header should be put into this cluster. In other words, the cluster is broken. In our scenario, the sensors are the same type. There may be a sensor under the shadow and its neighbor sensor is under sunlight, or probably one or more sensors is damaged in that moment and  reports incorrect data. Keep in mind, in accordance with the principle of data clustering, irrelevant data is not located in the cluster.
 \begin{figure*}[htbp!]
 \centering
        \includegraphics[width=5in,height=10.5cm]{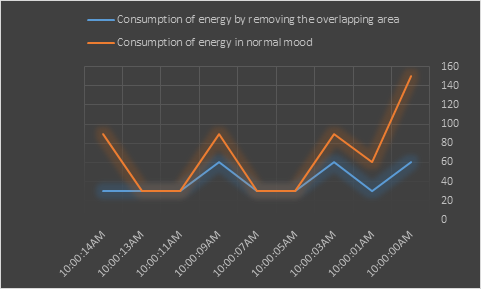}
        \centering
        \caption{Comparison of energy consumption in the normal state with the case where the overlapping area of clusters is eliminated.}
        \label{Graph D}
        
\end{figure*}
  To improve the work of [10], we changed the index tree   suggested by the  authors, and when the error data is sent instead of correct data to the base station, we suggest using the new tree structure presented in Figure 2.
\section{Evaluation}

We implemented a program based on our solution and achieved the following results. The number of nodes did not affect our experiment.  
\begin{table}[htbp]
\centering
  \caption{Comparison of the difference between the cluster head data and the real data of a node.}
  \resizebox{0.5\textwidth}{!}{
  \begin{tabular}{|c|c|}
     \hline
      Number of clusters & The difference between the cluster's head and the actual data \\ \hline
    1 & 1 \\ \hline
    2 & 2 \\ \hline
    3 &   1 \\ \hline
    4 & 0.5 \\ \hline
  \end{tabular}
  }
\end{table}

We achieved the above results in the implementation of the program and we concluded that increasing  the number of headers increases the accuracy of data sent to the base station.

In the article we chose, there are heterogeneous clusters, which means there is a lot of difference in the two ends of a cluster. Data filtering, especially for the averaging function, does not pass correct data from data of a cluster and in this case, we need to add a few headers to the cluster and transfer information to the base station with several headers instead of one header. This means that large clusters should be broken into several clusters. We performed this manner with K-Means algorithm and implemented this algorithm in C ++. We measured energy consumption in normal mode as well as the state in which overlapping cluster areas  were removed. We did this work by writing a program in Java environment and we got the following results.
\begin{table}[htbp]
  \centering
  \caption{  Comparison of energy consumption in the normal state with the case where the overlapping area is eliminated.}
  \begin{small}
  \tiny
\begin{tabular}{|p{0.72cm}|p{0.44cm}|p{0.39cm}|p{0.39cm}|p{0.39cm}|p{0.39cm}|p{0.39cm}|p{0.39cm}|p{0.39cm}|p{0.39cm}|p{0.39cm}}
  \hline
       &\begin{sideways}10:00:00 AM \ \ \end{sideways}&\begin{sideways}10:00:01 AM\end{sideways}&\begin{sideways}10:00:03 AM\end{sideways}&\begin{sideways}10:00:05 AM\end{sideways}&\begin{sideways}10:00:07 AM\end{sideways}&\begin{sideways}10:00:09 AM\end{sideways}&\begin{sideways}10:00:11 AM\end{sideways}&\begin{sideways}10:00:13 AM\end{sideways}&\begin{sideways}10: 00:14 AM\end{sideways}\\ \hline
   Consumption of energy by removing the overlapping area & 60  &30 & 60& 30 &30&60&30&30&30\\ \hline
    Consumption of energy in normal mood&150&60&90&30&30&90&30&30&90\\ \hline
  \end{tabular}
  
\end{small}
\end{table}
\section{Conclusion}
We were centralized on [10] that focused on grid-based clustering and the construction of an index tree. Data filtering, especially in the case of the average aggregation cannot transfer correct data from  the data of one cluster. On these occasions, we had to add a few head nodes in clusters and data transmitted to the base station by a few head nodes; this means that clusters should be broken into a few large clusters. We did this with the K-Means algorithm and in C++ implementation and  we tried to improve its clustering and made a change in its proposed index tree. On the other hand, it happens very often that repetitive information is sent by the two overlapping clusters to the base station and we were able to save energy by removing duplicate data. We are clustering in the clusterable network area so that when head clustering crashes, we prevent the network from disconnecting by adding some new sensors making it easy and energy saving.

\begin{IEEEbiographynophoto}{Arezoo Khatibi}

 correspond email: \protect\url{arezoo.khatibi@grad.kashanu.ac.ir}, Faculty of Computer Science,  University of Kashan,  BLVD Ghotb Ravandi 6 kilometers, Kashan Iran.
 [{\includegraphics[width=1in,height=1.25in,clip,keepaspectratio]{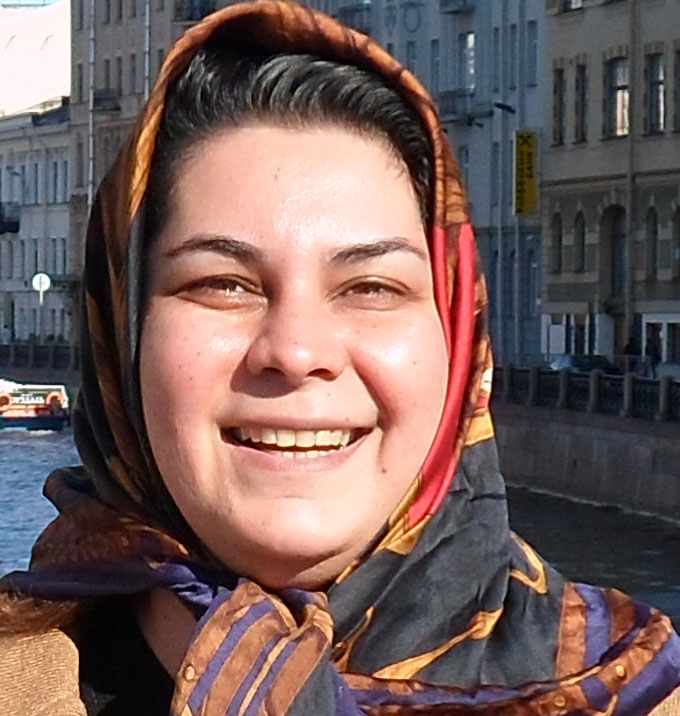}}]{Arezoo Khatibi}

\end{IEEEbiographynophoto}
\end{document}